\documentclass[twoside,twocolumn,9pt]{article}
\usepackage{extsizes}
\usepackage[super,sort&compress,comma]{natbib} 
\usepackage[version=3]{mhchem}
\usepackage[left=1.5cm, right=1.5cm, top=1.785cm, bottom=2.0cm]{geometry}
\usepackage{balance}
\usepackage{sectsty}
\usepackage{graphicx} 
\usepackage{lastpage}
\usepackage[format=plain,justification=justified,singlelinecheck=false,font={stretch=1.125,small,sf},labelfont=bf,labelsep=space]{caption}
\usepackage{float}
\usepackage{fancyhdr}
\usepackage{fnpos}
\usepackage[english]{babel}
\addto{\captionsenglish}{%
  
}
\usepackage{array}
\usepackage{droidsans}
\usepackage{charter}
\usepackage[T1]{fontenc}
\usepackage[usenames,dvipsnames]{xcolor}
\usepackage{setspace}
\usepackage[compact]{titlesec}
\usepackage{hyperref}
\usepackage{amssymb}
\usepackage{amsmath}
\usepackage{bm}
\usepackage{mathrsfs}
\usepackage{tcolorbox}
\usepackage{natbib}
\usepackage{bibentry}
\usepackage{mathrsfs}
\usepackage{relsize}
\usepackage{array}
\usepackage{booktabs}
\newcommand{\ket}[1]{| #1 \rangle}

\newcommand{\eqr}[1]{Eq.~\eqref{eq:#1}}

\newcommand{\eql}[1]{\label{eq:#1}}
\newcommand{\figr}[1]{Fig.~\ref{fig:#1}}
\newcommand{\figl}[1]{\label{fig:#1}}
\newcommand{\bse}{\begin{subequations}}
\newcommand{\ese}{\end{subequations}}

\newcommand{\tabr}[1]{Table~\ref{tab:#1}}
\newcommand{\tabl}[1]{\label{tab:#1}}

\renewcommand{\thefootnote}{\alph{footnote}}

\definecolor{cream}{RGB}{222,217,201}

\begin{document}

\pagestyle{fancy}
\thispagestyle{plain}
\fancypagestyle{plain}{
%%%HEADER%%%
\renewcommand{\headrulewidth}{0pt}
}
%%%END OF HEADER%%%

%%%PAGE SETUP - Please do not change any commands within this section%%%
\makeFNbottom
\makeatletter
\renewcommand\LARGE{\@setfontsize\LARGE{15pt}{17}}
\renewcommand\Large{\@setfontsize\Large{12pt}{14}}
\renewcommand\large{\@setfontsize\large{10pt}{12}}
\renewcommand\footnotesize{\@setfontsize\footnotesize{7pt}{10}}
\makeatother

\renewcommand{\thefootnote}{\fnsymbol{footnote}}
\renewcommand\footnoterule{\vspace*{1pt}% 
\color{cream}\hrule width 3.5in height 0.4pt \color{black}\vspace*{5pt}} 
\setcounter{secnumdepth}{5}

\makeatletter 
\renewcommand\@biblabel[1]{#1}            
\renewcommand\@makefntext[1]% 
{\noindent\makebox[0pt][r]{\@thefnmark\,}#1}
\makeatother 
\renewcommand{\figurename}{\small{Fig.}~}
\sectionfont{\sffamily\Large}
\subsectionfont{\normalsize}
\subsubsectionfont{\bf}
\setstretch{1.125} %In particular, please do not alter this line.
\setlength{\skip\footins}{0.8cm}
\setlength{\footnotesep}{0.25cm}
\setlength{\jot}{10pt}
\titlespacing*{\section}{0pt}{4pt}{4pt}
\titlespacing*{\subsection}{0pt}{15pt}{1pt}
%%%END OF PAGE SETUP%%%

%%%FOOTER%%%
\fancyfoot{}
\fancyfoot[LO,RE]{\vspace{-7.1pt}\includegraphics[height=9pt]{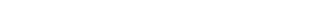}}
\fancyfoot[CO]{\vspace{-7.1pt}\hspace{13.2cm}\includegraphics{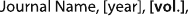}}
\fancyfoot[CE]{\vspace{-7.2pt}\hspace{-14.2cm}\includegraphics{head_foot/RF}}
\fancyfoot[RO]{\footnotesize{\sffamily{1--\pageref{LastPage} ~\textbar  \hspace{2pt}\thepage}}}
\fancyfoot[LE]{\footnotesize{\sffamily{\thepage~\textbar\hspace{3.45cm} 1--\pageref{LastPage}}}}
\fancyhead{}
\renewcommand{\headrulewidth}{0pt} 
\renewcommand{\footrulewidth}{0pt}
\setlength{\arrayrulewidth}{1pt}
\setlength{\columnsep}{6.5mm}
\setlength\bibsep{1pt}
%%%END OF FOOTER%%%

%%%FIGURE SETUP - please do not change any commands within this section%%%
\makeatletter 
\newlength{\figrulesep} 
\setlength{\figrulesep}{0.5\textfloatsep} 

\newcommand{\topfigrule}{\vspace*{-1pt}% 
\noindent{\color{cream}\rule[-\figrulesep]{\columnwidth}{1.5pt}} }

\newcommand{\botfigrule}{\vspace*{-2pt}% 
\noindent{\color{cream}\rule[\figrulesep]{\columnwidth}{1.5pt}} }

\newcommand{\dblfigrule}{\vspace*{-1pt}% 
\noindent{\color{cream}\rule[-\figrulesep]{\textwidth}{1.5pt}} }

\makeatother
%%%END OF FIGURE SETUP%%%

%%%TITLE, AUTHORS AND ABSTRACT%%%
\twocolumn[
  \begin{@twocolumnfalse}
{\includegraphics[height=30pt]{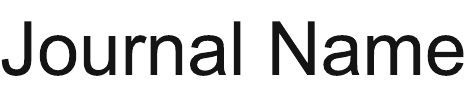}\hfill\raisebox{0pt}[0pt][0pt]{\includegraphics[height=55pt]{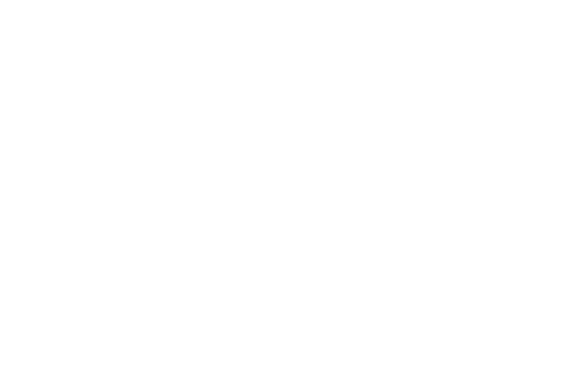}}\\[1ex]
\includegraphics[width=18.5cm]{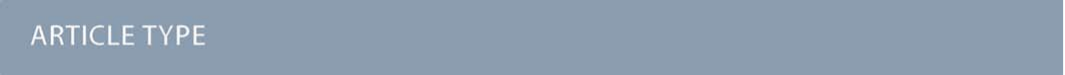}}\par
\vspace{1em}
\sffamily
\begin{tabular}{m{4.5cm} p{13.5cm} }

\includegraphics{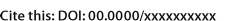} & \noindent\LARGE{\textbf{Learning Radical Excited States from Sparse Data$^\dag$}} \\%Article title goes here instead of the text "This is the title"
\vspace{0.3cm} & \vspace{0.3cm} \\

 & \noindent\large{Jingkun\ Shen,\textit{$^{a,\ddag}$} Lucy\ E. Walker,\textit{$^{b,c,\ddag}$} Kevin\ Ma,\textit{$^{a,\ddag}$} James D.\ Green,\textit{$^{a,\ddag}$} Hugo\ Bronstein,\textit{$^{b,c}$} Keith\ T. Butler\textit{$^{a}$} and Timothy J.\ H.\ Hele$^{\ast}$\textit{$^{a}$} } \\%Author names go here instead of "Full name", etc.

\includegraphics{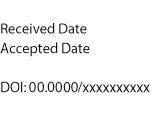} & \noindent\normalsize{Emissive organic radicals are currently of great interest for their potential use in the next generation of highly efficient organic light emitting diode (OLED) devices and as molecular qubits. However, simulating their optoelectronic properties is challenging, largely due to spin-contamination and the multireference character of their excited states. Here we present a data-driven approach where, for the first time, the excited electronic states of organic radicals are learned directly from experimental excited state data, using a much smaller amount of data than typically required by Machine Learning. We adopt ExROPPP, a fast and spin-pure semiempirical method for calculation of the excited states of radicals, as a surrogate physical model for which we learn the optimal set of parameters. To achieve this we compile the largest known database of organic radical geometries and their UV-vis data, which we use to train our model.
Our trained model gives Root Mean Square (RMS) and mean absolute errors for excited state energies of 0.24 and 0.16 eV respectively, improving hugely over ExROPPP with literature parameters. Four new organic radicals are synthesised and we test the model on their spectra, finding even lower errors and similar correlation as for the testing set.
This model paves the way for the high throughput discovery of next generation radical-based optoelectronics.} \\

\end{tabular}

 \end{@twocolumnfalse} \vspace{0.6cm}

  ]
%%%END OF TITLE, AUTHORS AND ABSTRACT%%%

%%%FONT SETUP - please do not change any commands within this section
\renewcommand*\rmdefault{bch}\normalfont\upshape
\rmfamily
\section*{}
\vspace{-1cm}

%%%FOOTNOTES%%%

\footnotetext{\textit{$^{a}$~Department of Chemistry, Christopher Ingold Building, University College London, WC1H 0AJ, UK.}}
\footnotetext{\textit{$^{*}$E-mail: t.hele@ucl.ac.uk}}
\footnotetext{\textit{$^{b}$~Yusuf Hamied Department of Chemistry, University of Cambridge, CB2 1EW, UK. }}
\footnotetext{\textit{$^{c}$~Department of Physics, Cavendish Laboratory, Cambridge University, Cambridge CB3 0HF, UK. }}
\footnotetext{\textit{$^{\ddag}$These authors contributed equally.}}
%Please use \dag to cite the ESI in the main text of the article.
%If you article does not have ESI please remove the the \dag symbol from the title and the footnotetext below. 
\footnotetext{\dag~Supplementary Information available: [details of any supplementary information available should be included here]. See DOI: 00.0000/00000000.}
%additional addresses can be cited as above using the lower-case letters, c, d, e... If all authors are from the same address, no letter is required

\date{\today}

\bibliographystyle{tim}

\section{Introduction}
\label{sec:intro}
%\subsection{Radical-based OLEDs}
Recent years have shown a great interest in radicals for organic light emitting diodes (OLEDs), which display internal quantum efficiencies (IQE) of near 100\% and intense emission in the deep red, NIR and IR spectral regions, features which are unusual and highly desirable.\cite{pen15a,ai18a,guo14a,bla02a,mur23a,abd20a,li22a,cho24a} These radical OLEDs, based on organic monoradicals, offer an alternative for the next generation of highly efficient lighting. Furthermore, the optical readout of the quartet state of some radicals has potential applications in quantum information science and paves the way for next-generation molecular qubits.\cite{gor23a} However, historically many organic radicals have been non-emissive, such that trial-and-error exploration of chemical space is inefficient, and there is therefore a large and unmet need for a method which facilitates the fast, accurate and spin-pure calculation of the low-lying excited states of a wide variety of radical molecules. Such a computational method would also be invaluable for the  high-throughput screening of radicals for their UV-visible spectra. This work focuses on organic monoradicals, i.e. molecules with only one unpaired electron, however, it should be noted that the excited state properties of organic biradicals and organic radicals with many unpaired electrons are generally different to those discussed here.\cite{yu24a,miz24b}

Calculating the excited electronic states of radicals is challenging due to spin-contamination and their multiconfigurational character. There exist several highly accurate methods for calculation of the excited states of radicals, such as multiconfigurational self-consistent field (MCSCF), complete active space perturbation theory to the 2\textsuperscript{nd} order (CASPT2) and Coupled-cluster theory, however, these methods are very computationally expensive, making them unsuitable for high-throughput workflows.\cite{and92a,nak93a,pur82a} Moreover, computationally cheaper methods such as conventional Time-Dependent Density Functional Theory (TD-DFT) can lead to spin-contaminated and functional-dependent results for the electronically excited states of radicals.\cite{he19a,li16a} Additionally, it has been shown that for the most accurate calculation of excited state energies one must also include nuclear quantum effects.\cite{hel21b} Recently, an alternative, semiempirical method was developed --- ExROPPP (Extended Restricted Open-shell Pariser-Parr-Pople theory), which is significantly faster, yet as accurate as higher level methods for calculating excited states of hydrocarbon radicals.\cite{gre24a} ExROPPP is a based on the Pariser-Parr-Pople (PPP) Hamiltonian\cite{par53a,par56a,pop53a,pop54a,mat57a} with a subsequent Extended Configuration Interaction Singles (XCIS)\cite{mau96a} calculation which ensures spin purity. PPP theory has gained recent popularity for predicting electronic properties at a significantly reduced computational cost.\cite{hel19a,gre22a,bed23a,gre24a,dub24a,jor24a} Being a semiempirical method, PPP theory and consequently ExROPPP requires parameters which must be specified at the start of a calculation.\cite{par53a,par56a,pop53a,pop54a,mat57a}  The carbon atom PPP parameters already existing in the literature have been shown to be successful for predicting excited state energies for hydrocarbons in ExROPPP.\cite{gre24a} However, emissive radicals commonly contain nitrogen and chlorine atoms, and we are not aware of any consistent, unified and widely-accepted set of parameters for including heteroatoms such as these in PPP theory or ExROPPP.\cite{mat57a,hin71a,van80a} The advent of ExROPPP has opened up the possibility of rapid screening of the electronically excited states of radicals, however, extending and generalizing this method requires an optimal set of parameters to be found.

In recent years Machine Learning (ML) has become an indispensable tool for the study of chemical systems.\cite{but18a} Such models allow for accurate prediction of chemical and physical properties with huge computational savings compared to methods such as DFT, provided sufficient data are available, and are often seen as an alternative to semiempirical and classical force-field methods.\cite{mon13a} ML has seen numerous applications in predicting energies, structures and reactivity patterns of molecules and materials.\cite{but18a,mon13a,ben18a,bar17a,der19a,srs24a,che18a} Furthermore it has been applied to calculating the excited states of molecules and simulating excited state potential energy surfaces.\cite{mon13a,ram15a,roc20a,che22b,wes21a,dra21a,dra18a,xue20a} However, while a wealth of previous studies have been successful for closed-shell species, we find very few examples of ML for the electronically excited states of radicals. In recent work ML was applied to calculate the electronically excited states of radicals by training on closed-shell molecules, however to our knowledge, learning the excited states directly from excited state data of organic radicals themselves has yet to be attempted. \cite{ju24a} Furthermore, typical ML models, in which no strong priors about the system are assumed at the outset, generally require large amounts of data, e.g. the properties of thousands of molecules or more, in order to be successful\cite{but18a,mon13a,bar17a} and we are unaware of any such large datasets for radicals. In addition, predicting electronically excited states using ML is challenging as they are a largely non-local property which cannot in general be treated using atom-wise descriptors.\cite{wes21a}   %, unlike, for example, the total energy which can be broken down into individual atomistic contributions. 
%Several non-local descriptors exist, but have their own inherent challenges.\cite{wes21a}  
Moreover, the prediction of primary outputs of quantum chemistry such as the $N$-electron wavefunction (and thus the composition of excited electronic states) is a highly desirable feature of an ML model, yet few ML models can predict these quantities for excited states.\cite{wes21a} There has, however, been much recent interest in leveraging wavefunction-based descriptors in quantum ML models allowing them to retain some of the physical intuition of conventional quantum chemistry, and this work follows in a similar spirit to these developments.  \cite{fab22a,bri24a,wes20a}

Due to the lack of sufficient excited state data available in databases for organic radicals and the aforementioned challenges, adopting a trusted physical model such as ExROPPP and learning its optimal parameters may be a viable alternative to conventional ML for the excited states of radicals.\cite{hof63a,dew77a,ste07a,rep02a,chr16a,fab08a,hel19a,hel21a,gre22a,gre24a} Using such a model also allows for the direct prediction of a variety of primary quantum chemical quantities such as molecular orbitals and transition dipole moments. In this paper we will focus on predicting molecular UV-visible linear absorption spectra and leave the computation of emission spectra, which usually requires excited state geometries that are difficult to acquire, for future research. 

The linear UV-visible absorption spectra of organic radicals are usually characterised by two main features. These are an intense absorption (or absorptions) in the UV, usually between 300-400 nm, and a much weaker absorption in the visible. The weak visible $D_1$ state has been investigated using various levels of theory (TD-DFT, PPP, MCSCF).\cite{ai18a,gre24a,gor23a} In the special case of alternant hydrocarbons, $D_1$ is a minus combination $\ket{\Psi_{i0}^-}$ of the HOMO-SOMO and SOMO-LUMO excitations and is essentially dark in the absorption spectrum.\cite{chu54a,lon55a,abd20a,hud21a,gre24a} Alternant hydrocarbons are usually non-emissive as non-radiative processes outcompete fluorescence.\cite{abd20a,hel21b} Conversely, in non-alternant molecules, the $D_1$ state may have significantly higher absorption intensity. One widely explored class of non-alternant radicals are those with a donor-acceptor structure, such as TTM-1Cz, in which the $D_1$ state is bright, charge transfer (CT) in nature and is mostly composed of the HOMO-SOMO excitation.\cite{pen15a,ai18a,gor23a} These radicals are also highly emissive and have been incorporated into high-performing OLEDs.\cite{pen15a,ai18a} Another class of emissive radicals have recently been discovered which lack a donor-acceptor structure and CT characteristics, but instead employ mesityl groups leading to a large increase in the photoluminescence quantum yield. However, mesityl substitution does not significantly affect the absorption characteristics.\cite{mur23a} %In alternant hydrocarbons the intense UV absorptions are quite distinct and are usually made up of the states $\ket{\Psi_{i,0}^+}$ which are the in-phase combinations of single- excitations involving the SOMO.\cite{gre24a} However, it is harder to generalise about the origin of the bright UV absorptions in non-alternant organic radicals, and also as emission occurs from $D_1$, these higher-lying states have been less well studied than $D_1$.

In this paper we learn the excited states of organic radicals directly from their experimental data for the first time. To achieve this we use a modest amount of published UV-visible absorption data 
%of organic radicals containing carbon, hydrogen, nitrogen and chlorine and
to learn an optimal set of ExROPPP parameters for organic radicals containing carbon, hydrogen, nitrogen and chlorine. Despite only containing a modest amount of data, we believe our compiled database of UV-Visible absorption data of $\pi$-conjugated organic radicals to be the largest of its kind.\cite{ju24a}
%for these atoms. 
Four new radicals are synthesised and we test our model on their absorption spectra to demonstrate its predictability and transferability. We find that the trained model has a significantly higher accuracy than the model using parameters taken from the literature and is able to make accurate predictions about the electronic excited states of unseen molecules. 

\section{Methodology}
\label{sec:mthd}

\subsection{Data collection}
We obtained spectroscopic data for 81 organic radicals from previously published work whose structures are given in \figr{trn_mol1} and \figr{trn_mol2}.\cite{miz24a,mur23a,ju24a,ai18a,gor23a,abd20a,ai18b,rub93a,li22a,mat22a,gao17a,don17a,guo19a,don18a,yan22a} In order to compile a database of suitable radicals, we considered all radicals we could find in the literature containing only carbon, hydrogen, chlorine and pyrrole, aniline and pyridine type nitrogen atoms. Those radicals whose spectroscopic absorption data could be found were added to the database along with their data. We also obtained DFT optimised molecular geometries for these molecules from previous studies.\cite{ju24a,mur23a,gor23a,ai18a,ai18b} However, the molecular geometries for some molecules could not be found in the literature so these structures were optimised using unrestricted DFT in GAMESS-US.\cite{gamess} %Most molecules were optimized using the B3LYP functional, however, for a few PBE was instead used due to geometry convergence issues with B3LYP. The 6-31G(d,p) basis set was used for all geometry calculations. 
These data constitute the training set of the ML ExROPPP model. Further details pertaining to the collation of the database and geometry calculations can be found in SI Section I.
% \clearpage

\begin{figure*}
\centering
\includegraphics[width=0.9\textwidth]{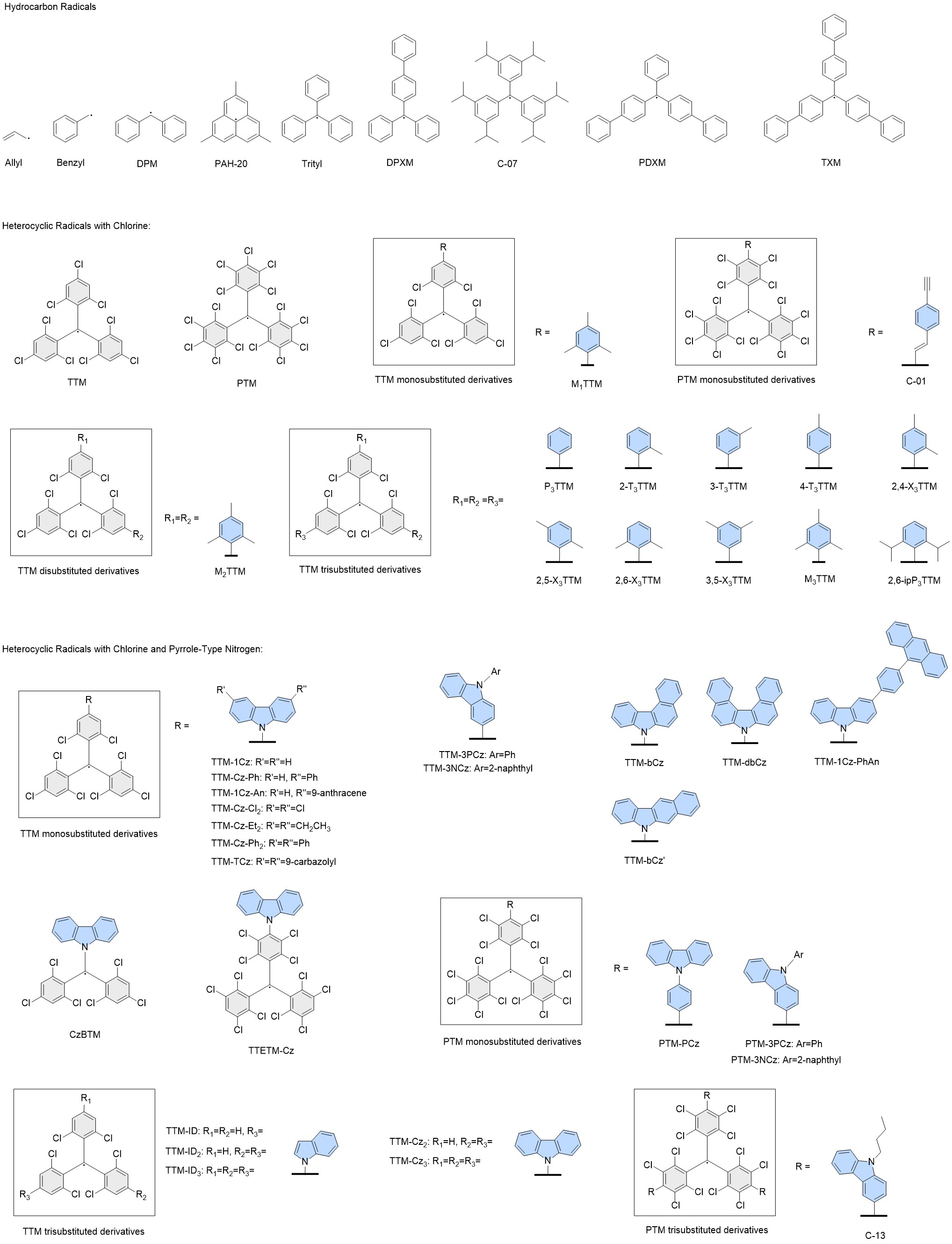}
\caption{Structures of the molecules in the training set containing carbon, hydrogen, chlorine and pyrrole-type nitrogen. The central TTM/PTM backbone is coloured in grey, and substituents colored in light blue.} %Grey color represents the conjugated parts in TTM/PTM, and light blue color shows the conjugated parts in the substitutes.}
\figl{trn_mol1}
\end{figure*}
% \clearpage
\begin{figure*}
\centering
\includegraphics[width=0.9\textwidth]{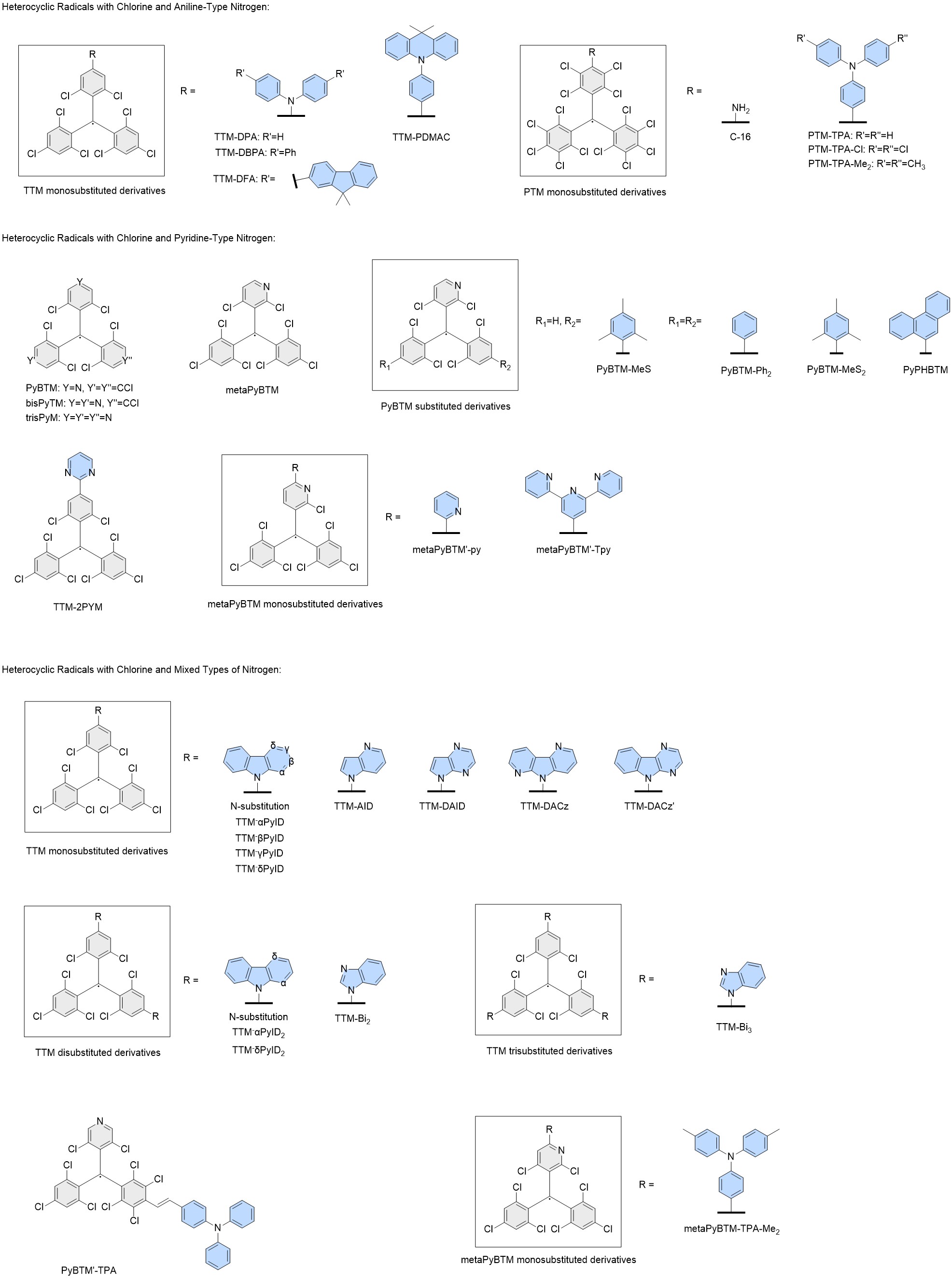}
\caption{Structures of the molecules in the training set containing aniline, pyridine and multiple types of nitrogen. The central TTM/PTM backbone is coloured in grey, and substituents colored in light blue.}
\figl{trn_mol2}
\end{figure*}
% \clearpage

The target properties used for training are the energies $E_{D_1}$ of the the first excited doublet states ($D_1$), the energies $E_\mathrm{brt}$ of the brightest absorptions in the UV-visible spectra, and $I_{D_1}^{\textrm{rel.}}=\epsilon_{D_1}/\epsilon_{\textrm{brt}}$ which is the ratio of the molar extinction coefficients of these the two absorptions, extracted from linear UV-visible absorption spectra. These target properties are similar to those previously used in ML of molecular spectra\cite{mon13a}. Exceptions are made for TTM-1Cz-An and TTM-1Cz-PhAn whereby due to their unusual electronic structure, their first excited doublet state ($D_1$) is a dark triplet-coupled doublet state and the lowest energy bright doublet state is $D_2$ which is by majority composed of the carbazole HOMO-TTM SOMO excitation (same orbital parentage as $D_1$ in typical radicals).\cite{gor23a} Therefore, for these exceptional molecules we fit the $D_2$ (instead of $D_1$) energy and oscillator strength to the corresponding lowest energy $D_2$ absorption seen in experiment. For these two molecules, we will group this state ($D_2$) in with the $D_1$ states for all the other molecules when performing the statistical analyses. 

\subsection{Training}
We train the ExROPPP model on experimental UV-visible data of known organic radicals, using a fitness function of the computed energies and intensities compared to those obtained from experiment, which quantifies how well the predictions of the ExROPPP model fit with the experimental data. The fitness function takes the form
\begin{align}
\eql{fit}
f =& w_{D_1}(E_{D_1,\mathrm{calc.}}-E_{D_1,\mathrm{exp.}})^2 + w_{\mathrm{brt}}(E_{\mathrm{brt},\mathrm{calc.}}-E_{\mathrm{brt},\mathrm{exp.}})^2 \nonumber \\
&+ w_I(I_{D_1,\mathrm{calc.}}^{\mathrm{rel.}}-I_{D_1,\mathrm{exp.}}^{\mathrm{rel.}})^2,
\end{align}
where $w_{D_1}$, $w_{\mathrm{brt}}$ and $w_I$ are the weights of the three respective terms in \eqr{fit}. The weights of the first two terms have units of $\mathrm{eV}^{-2}$ and $w_I$ is dimensionless such that $f$ is a dimensionless quantity. While there are many other fitness functions which we could use, such as those based on the theory of optimal transport (between experimental and calculated spectra), we choose to use the above function as it encapsulates the essential spectral information of organic radicals which we believe is most important to be able to predict and only requires a small amount of raw spectral data for each radical.\cite{sei21a} 

\begin{figure}[tb]
    \centering
    \includegraphics[width=0.5\textwidth]{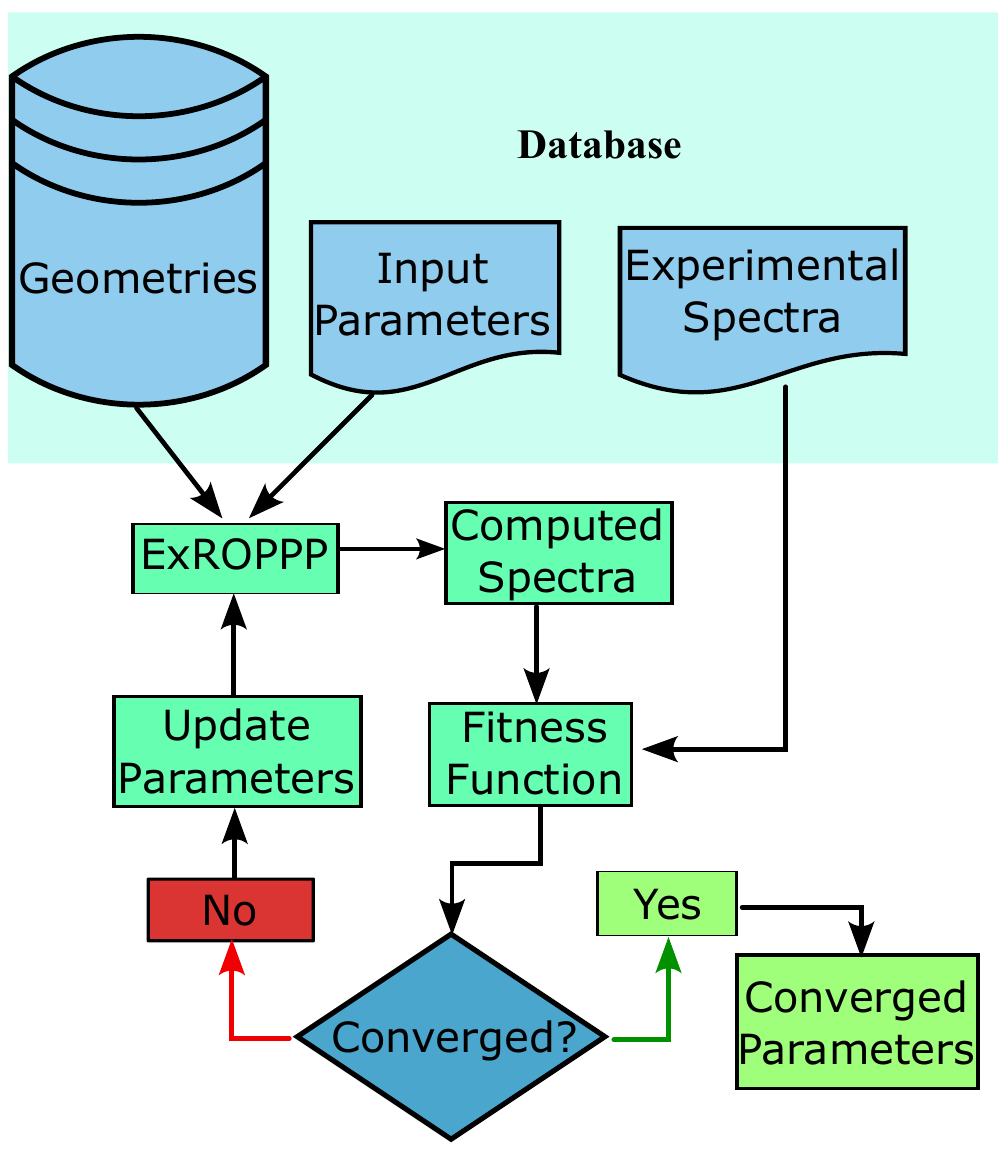}
    \caption{A flow diagram illustrating our method for training our ExROPPP model on the experimental absorption data of organic radicals.}
    \label{fig:kevin_flow}
\end{figure}

Training is achieved by finding a set of ExROPPP parameters which minimises this fitness function utilising the derivative-free Nelder-Mead optimiser in Python as shown in \figr{kevin_flow}.\cite{nel65a} The algorithm first reads in the initial parameters, molecular geometries and experimental absorption data for all training molecules and classifies the molecules into hydrocarbons or heterocycles, which are treated slightly differently. For hydrocarbons, the fitness function comprises of only energy terms, with $w_{D_1}$ and $w_{\mathrm{brt}}$ set to 1 and $w_I$ set to zero as the $D_1$ state for hydrocarbons gives a very weak absorption in experiment and in ExROPPP has zero oscillator strength.\cite{chu54a} For heterocycles, all three terms are included with weights of 1 (except for a few molecules whose bright state data could not be found, see radicals-spreadsheet.xlsx in Data Availability). Then the parameters are iteratively varied and the fitness calculated on each iteration until convergence. 

We found that pre-training separately on subsets of molecules of different heteroatom types to obtain better initial guess parameters before training on all molecules for all parameters, an approach which we call `stratified' training, lead to lower errors than not including these pre-training steps. The results of the trained model presented in the next section are obtained using this stratified approach. Details of the training process are discussed further in SI Section II. These calculations are parallelised for maximum efficiency.

% \begin{figure}[tb]
% \includegraphics[width=0.5\textwidth]{RSC ChemDraw/untitled.jpg}
% \caption{Structures of the four newly synthesised radicals reported in this work: M\textsubscript{2}TTM-4Me, M\textsubscript{2}TTM-3PCz, M\textsubscript{2}TTM-3TPA and M\textsubscript{2}TTM-4TPA, which constitute the testing set.}
% \figl{lucy_mol}
% \end{figure}
%\subsection{$K$-fold validation}

\subsection{Testing on novel radicals}
Four new radicals: M\textsubscript{2}TTM-4Me, M\textsubscript{2}TTM-3PCz, M\textsubscript{2}TTM-3TPA and M\textsubscript{2}TTM-4TPA, shown in \figr{lucy_mol} were synthesised, their UV-visible absorption spectra were measured, and their minimum energy geometries were obtained using DFT. The spectroscopic data ($E_{D_1}$, $E_\mathrm{brt}$ and $I_{D_1}^{\textrm{rel.}}$) and molecular structures of these new molecules form a testing set for the ML ExROPPP model. The molecular geometries and extracted UV-visible absorption data of all molecules as well as initial and optimised sets of parameters are available on the UCL Research Data Repository (see Data Availability). 
\subsection{Statistical analysis}
We calculate root mean-squared errors (RMSE), mean absolute differences (MAD), $\mathrm R^2$ and Spearman's rank correlation coefficients (SRCC) between the experimental data and the simulated data for the training and testing set (see SI Section IV C).
\subsection{PPP parameterization}
We employ largely the same functional form of PPP theory as in previous work. \cite{hel19a,gre22a,gre24a,mat57a} The parameters of this model are the one-electron on-site Coulomb $\epsilon_\mu$ and hopping $t$ parameters, and two-electron Hubbard $U$ and distance scaling $r_0$ parameters. We use the Mataga-Nishimoto form for the two-electron integrals
\begin{align}
 \gamma_{\mu\nu} \simeq & (\mu\mu|\nu\nu) \nonumber\\
 = & \frac{U_{\mu \nu}}{1 + r_{\mu \nu}/r_{0,\mu \nu}}, 
\end{align}
expressed in terms of atomic orbitals, where $r_{\mu \nu}$ is the scalar distance between atoms $\mu$ and $\nu$.\cite{mat57a}
However unlike Refs.~\citenum{hel19a,gre22a,gre24a,mat57a}, but similar to Ref.~\citenum{bru71a} we elect to use an exponentially decaying function which is scaled by the cosine of the dihedral angle for the hopping term of the form $t_{\mu \nu}=A\exp(-br_{\mu \nu})\cos\theta$. The PPP parameters are atom specific, with a different $\epsilon_\mu$ for each atom type and different $t_{\mu \nu}$ for each pair of types of bonded atoms $\mu$ and $\nu$. There is only one independent $U_{\mu \mu}$ and $r_{0,\mu \mu}$ for each atom type, and an average of the parameters for different atom types is taken for two-electron interactions between two different types of atoms i.e. $U_{\mu \nu}=\frac{1}{2}(U_{\mu \mu}+U_{\nu \nu})$ and $r_{0,\mu \nu}=\frac{1}{2}(r_{0,\mu \mu}+r_{0,\nu \nu})$. We use different parameters for pyridine and pyrrole/aniline type nitrogen atoms due to their different numbers of $\pi$-electrons. Carbon and pyridine type nitrogen atoms contribute one electron to the $\pi$-system whereas chlorine and pyrrole/aniline type nitrogen atoms contribute two, from lone pairs. We model the atomic cores formed of the nuclei and core electrons as point charges at the nuclei with effective charge of $e$ for carbon and pyridine type nitrogen, and $2e$ for chlorine and pyrrole/aniline type nitrogen, where $e$ is the electron charge.    

To show the improvement of the trained model, we compare the trained parameters with those initially sourced from the literature. Full details of the parameterisation used can be found in SI Section V.

\section{Results}
\label{sec:result} 

\subsection{Training and Validation}
The results of training the 81 molecule model are summarized in \figr{comp_plot_train} and \tabr{stats_train}. We find that the accuracy of the simulated excited state energies improves significantly on training. The RMSE reduces from 0.86 eV with literature parameters to 0.24 eV for the trained model and the MAD reduces from 0.80 eV with literature parameters to 0.16 eV for the trained model. In terms of correlation, we find a marked improvement in $\mathrm{R}^2$ from $-$0.71 to 0.87 and a smaller improvement for SRCC going from 0.79 to 0.88 in the trained model compared with the literature parameters.

\begin{figure*}[tb]
\centering
\includegraphics[width=0.8\textwidth]{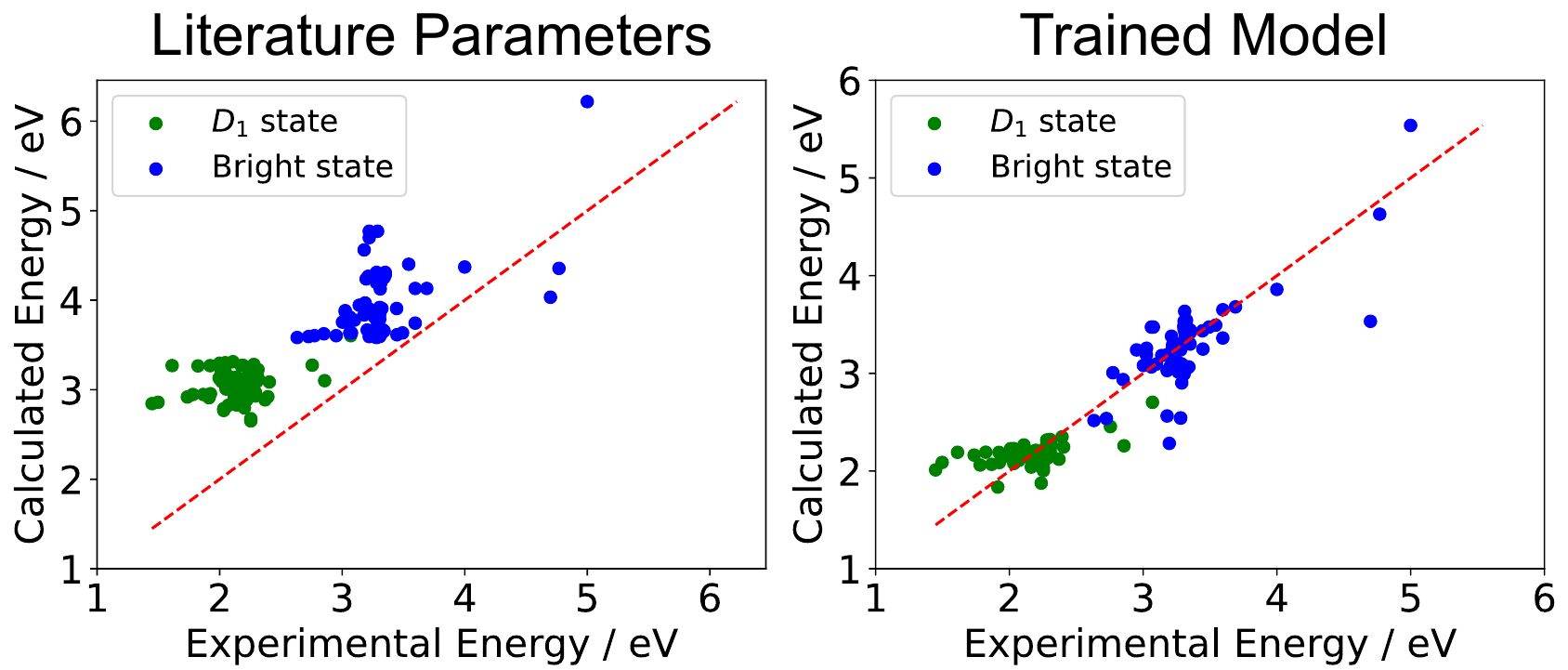}
\caption{Regression plots of the excited state energies of the 81 molecule training set calculated by ExROPPP compared with experimentally determined energies, using parameters obtained from the literature (left) and those of the trained ExROPPP model (right). The trained model predicts the energies of UV/Visible absorptions much closer to experiment (red line) than do the literature parameters.}
\figl{comp_plot_train}
\end{figure*}

\begin{figure*}[tb]
\centering
\includegraphics[width=0.75\textwidth]{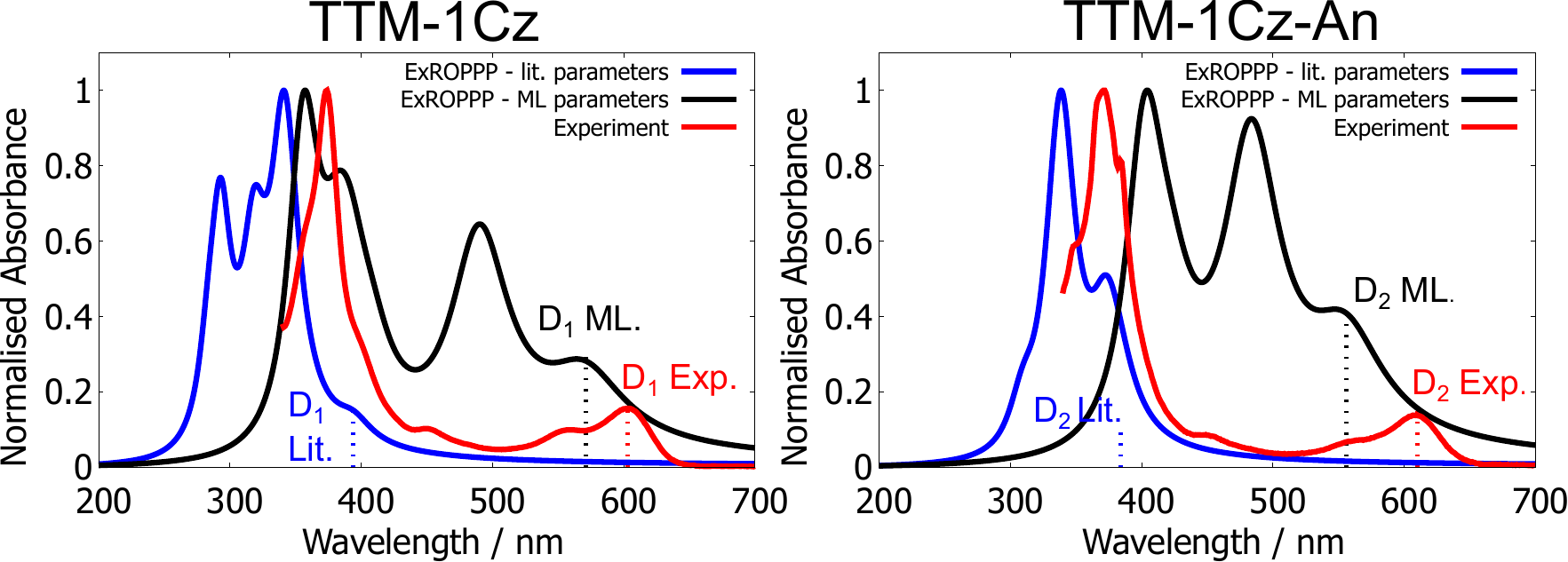}
\caption{UV-visible absorption spectra of TTM-1Cz and TTM-1Cz-An measured in 200 $\mathrm \mu $M toluene solution at room temperature (red), simulated using ExROPPP with literature parameters (blue) and with the trained 81 molecule model (black).\cite{gor23a} The trained model substantially improves on the literature parameters in both cases.}
\figl{abs_train}
\end{figure*}
The simulated spectra of the training set are more accurately reproduced by the trained model than by the model with literature parameters. To illustrate this we have included the spectra of two emissive radicals in the training set which are relevant to optoelectronics: TTM-1Cz and TTM-1Cz-An (see \figr{abs_train}). TTM-1Cz is a prototypical and widely-studied emissive radical which has been implemented in functioning OLEDs, and is a good reference point for the trained ExROPPP model.\cite{pen15a} On the other hand, TTM-1Cz-An is an atypical organic radical which has a complex and unusual electronic structure owing to its first excited state being a quartet and should be a challenging test case for ExROPPP. TTM-1Cz-An has been investigated for potential applications in quantum information technology.\cite{gor23a} We find that the trained model reproduces the $D_1$ ($D_2$ for Cz-An) energies of these molecules significantly more accurately than does the literature parameters. The accuracy for the bright states also improves with the trained parameters. Furthermore, ExROPPP predicts that the quartet state of TTM-1Cz-An is lower in energy than the lowest energy bright state $D_2$ in both sets of parameters, in line with experimental data and higher-level calculations in the literature.\cite{gor23a} Trained ExROPPP does produce an extra absorption around 500nm for both molecules which is not seen in experiment, and could be an artifact of the choice of fitness function. The fitness only depends on the energies and relative intensities of the $D_1$ ($D_2$ for Cz-An) state and the intense bright state and thus does not take into account other states in the spectrum. Nevertheless, the ability of the trained ExROPPP model to accurately capture the absorption spectra and excited state features of both typical and anomalous radicals shows its flexibility and robustness. 
 
\begin{table}
\caption{Total fitness, root mean-squared errors (RMSE), mean absolute differences (MAD), $\mathrm R^2$ and Spearman's rank correlation coefficients (SRCC) for the trained ExROPPP model and from K-fold validation compared to ExROPPP with parameters obtained from the literature, calculated for all states in the training set of 81 organic radicals. For further details on K-fold validation see SI section III B.}
% \begin{tabular}{c|c|c|c|c}
% &Literature& Trained & K-fold & Target\\ 
% &parameters & Model & Model & \\ \hline 
% Total Fitness &117.44 &10.00 & 9.70 &\\
% RMSE (all states)/eV &0.86 & 0.24 & 0.24 & $<0.3$\\
% MAD (all states)/eV &0.80 & 0.16& 0.16 & $<0.3$\\
% $\mathrm R^2$ (all states) & $-$0.71& 0.87 & 0.87& close to 1\\
% SRCC (all states) & 0.79& 0.88 & 0.87& close to 1
% \end{tabular}
% \tabl{stats_train}
% \end{table}

\begin{tabular}{c|c|c|c|c}
&Literature& Trained & K-fold & Target\\ 
&parameters & Model & Model & \\ \hline 
Total Fitness &117.44 &10.00 & 15.21 &\\
RMSE &0.86 & 0.24 & 0.27 & $<0.3$\\
(all states)/eV & &  &  & \\
MAD &0.80 & 0.16& 0.18 & $<0.3$\\
(all states)/eV & & &  &  \\
$\mathrm R^2$ (all states) & $-$0.71& 0.87 & 0.84& close to 1\\
SRCC & 0.79& 0.88 & 0.86& close to 1\\
(all states) & & & & 
\end{tabular}
\tabl{stats_train}
\end{table}

%To assess the predictivity and stability of the trained model, we employed a $K$-fold cross validation on the training set with 9 folds (see SI). As can be seen from Table S1 in the supplementary information, the errors for each of the 9 folds are extremely similar and with a similar error to the full training set (RMSE for $K$ folds in the range of \textbf{correct this with jingkun new k-fold} \textcolor{red}{0.23}--\textcolor{red}{0.24} eV, RMSE for training set is 0.24 eV). This shows that the trained model is predictive within the training set and excluding any one molecule or a small number of molecules from the training set does not affect the model. Furthermore, this is evidence that overfitting did not occur.
In addition, we employed $K$-fold cross-validation on the 81-radical training set to validate the robustness of the model against data noise, using several different choices of folds. %The results showed that for all methods of $K$-fold validation, 
The data noise is shown to be small for all trials. Artificially (selected by the user) and randomly distributed folds gave similar errors 
%(9-fold average RMSE differed by $\simeq12\%$) 
proving that the distribution of molecules across folds does not significantly affect the model. Also, the $K$-fold results show that the stratified model has better robustness than the model without pre-training steps.
%We averaged the results from artificially and randomly distributed folds, and 
%The standard deviations of the fitness, RMSE, MAD, R$^2$ and SRCC of all 18 folds (artificially and randomly distributed) are reported alongside the metrics for the full 81-molecule trained model in \tabr{stats_train}. 
Further details can be found in the SI Section III.

\subsection{Transferability}
Here we briefly consider the extent to which the trained ExROPPP model can reproduce the qualitative orbital structure of radicals. This is a particularly challenging test given that ExROPPP was not trained on orbital data.

A qualitative comparison of the singly occupied (SOMO) and highest occupied (HOMO) molecular orbitals from ExROPPP and ROHF (B3LYP/6-31G(d,p)) demonstrates good overall agreement in both shape and localization, highlighting the transferability of the ExROPPP‐based parametrisation. In TTM‐1Cz, TTM‐3NCz, and M$_3$TTM, the SOMOs remain largely centered on the triarylmethyl carbon, with moderate extension onto the phenyl rings, whereas the HOMOs for TTM‐1Cz and TTM‐3NCz localize more strongly on the carbazole substituent. Notably, the local/global symmetry (e.g., $C_2$ for TTM‐1Cz, $C_3$ for M$_3$TTM) and corresponding irreps are preserved. These observations underscore the physical interpretability and transferability of parameters derived via an optimizer‐assisted physics-informed model, which can extrapolate untrained properties. Minor discrepancies occur for the M$_3$TTM HOMO, where ExROPPP underestimates the small orbital coefficients in the outer ring. Further details and figures illustrating these points are provided in SI Section IV E.

\subsection{Novel organic radicals}
\begin{figure}[tb]
\includegraphics[width=0.5\textwidth]{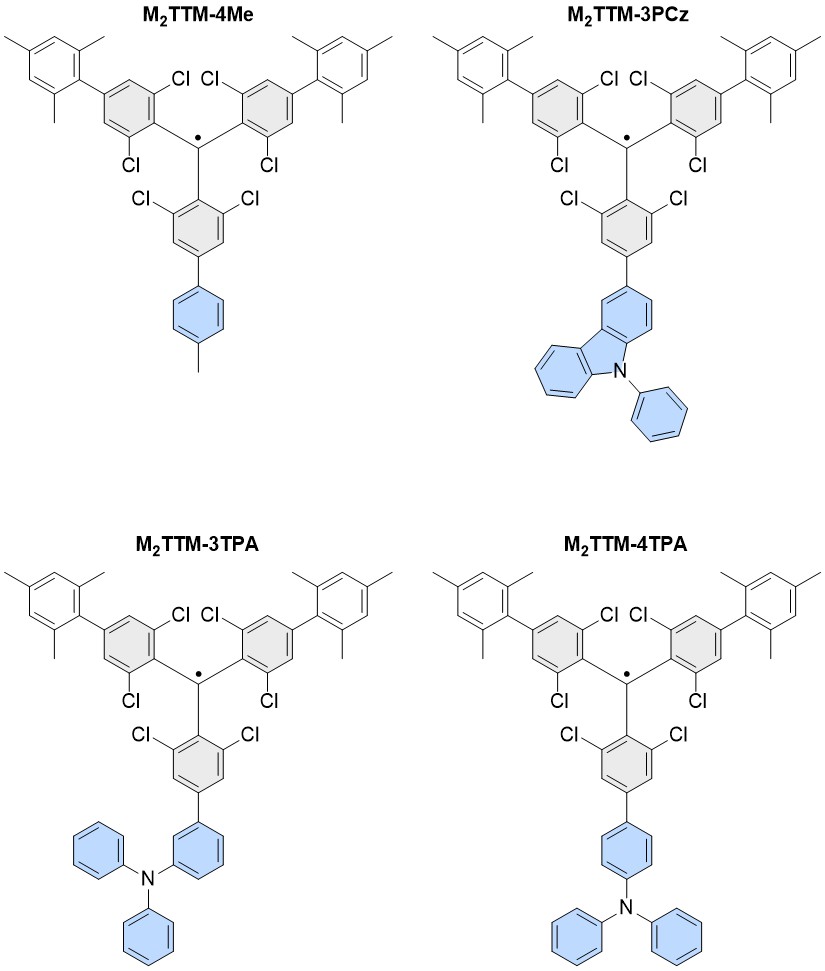}
\caption{Structures of the four newly synthesised radicals reported in this work: M\textsubscript{2}TTM-4Me, M\textsubscript{2}TTM-3PCz, M\textsubscript{2}TTM-3TPA and M\textsubscript{2}TTM-4TPA, which constitute the testing set.}
\figl{lucy_mol}
\end{figure}
To test our model, we synthesized four novel trityl radicals, specifically designed to probe various state-of-the-art concepts previously identified in mono-radical systems (see \figr{lucy_mol}). Each radical was based on a mesitylated TTM framework, which has been shown to enhance photoluminescence quantum efficiency (PLQE) by augmenting the radiative decay rate.\cite{mur23a} To evaluate the ExROPPP model with an asymmetric structure and the absence of charge transfer (CT), toluene was appended to the unsubstituted site of the mesitylated trityl radical core through its 4-position. The three other radicals incorporated CT groups, namely 9-phenylcarbazole (PCz) and triphenylamine (TPA), which contain non-bonding nitrogen lone pairs. These non-bonding electrons have been shown to enhance photoluminescent efficiency through a reduction in excitonic coupling to high-frequency vibrational modes.\cite{gho24a} Through the inclusion of PCz and TPA moieties, we aimed to test the model across electron-donating groups of varying strengths, with TPA being the stronger donor due to the hybridization of its nitrogen heteroatom which influences lone pair availability. Additionally, TPA units were linked to the trityl radical core through both the 3- and 4-positions to assess accuracy in predicting spectroscopic outcomes for different stereoisomers. The combination of mesitylation and non-bonding CT groups provides a promising strategy for developing highly efficient radical emitters and it is crucial that the ExROPPP method can predict outcomes for these cutting-edge radical designs. 
\begin{figure}
\centering
\includegraphics[width=0.3\textwidth]{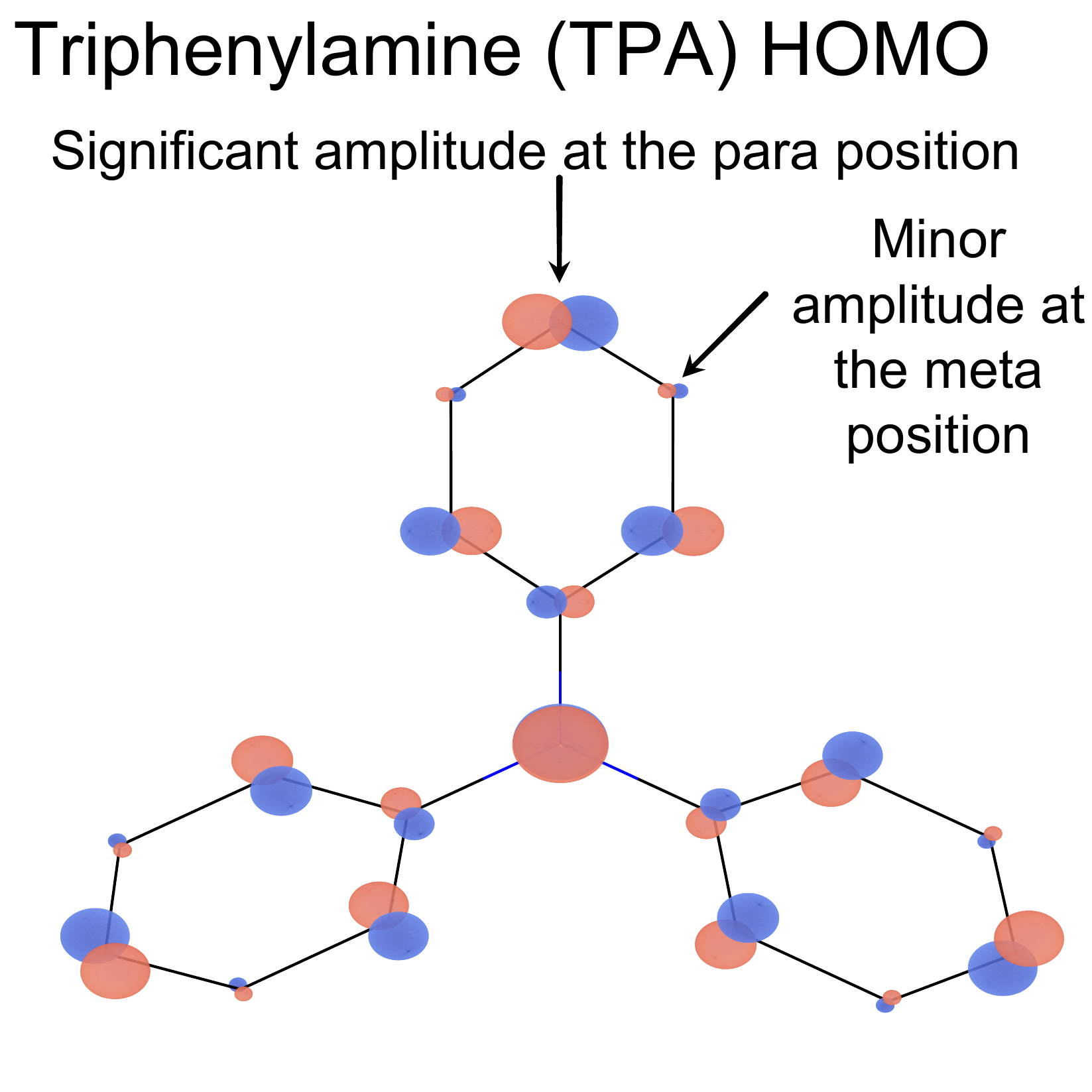}
\caption{HOMO of TPA calculated by closed-shell PPP (with the optimised parameters obtained from training on 81 radicals). There is significant HOMO amplitude at the para (4) position but minimal amplitude at the meta (3) position, such that the design rules correctly predict M$_2$TTM-4TPA to have a significant low-energy visible absorption and M$_2$TTM-3TPA not to have one.} 
\label{fig:tpa_homo}
\end{figure}

\begin{figure*}[h]
\centering
\includegraphics[width=0.8\textwidth]{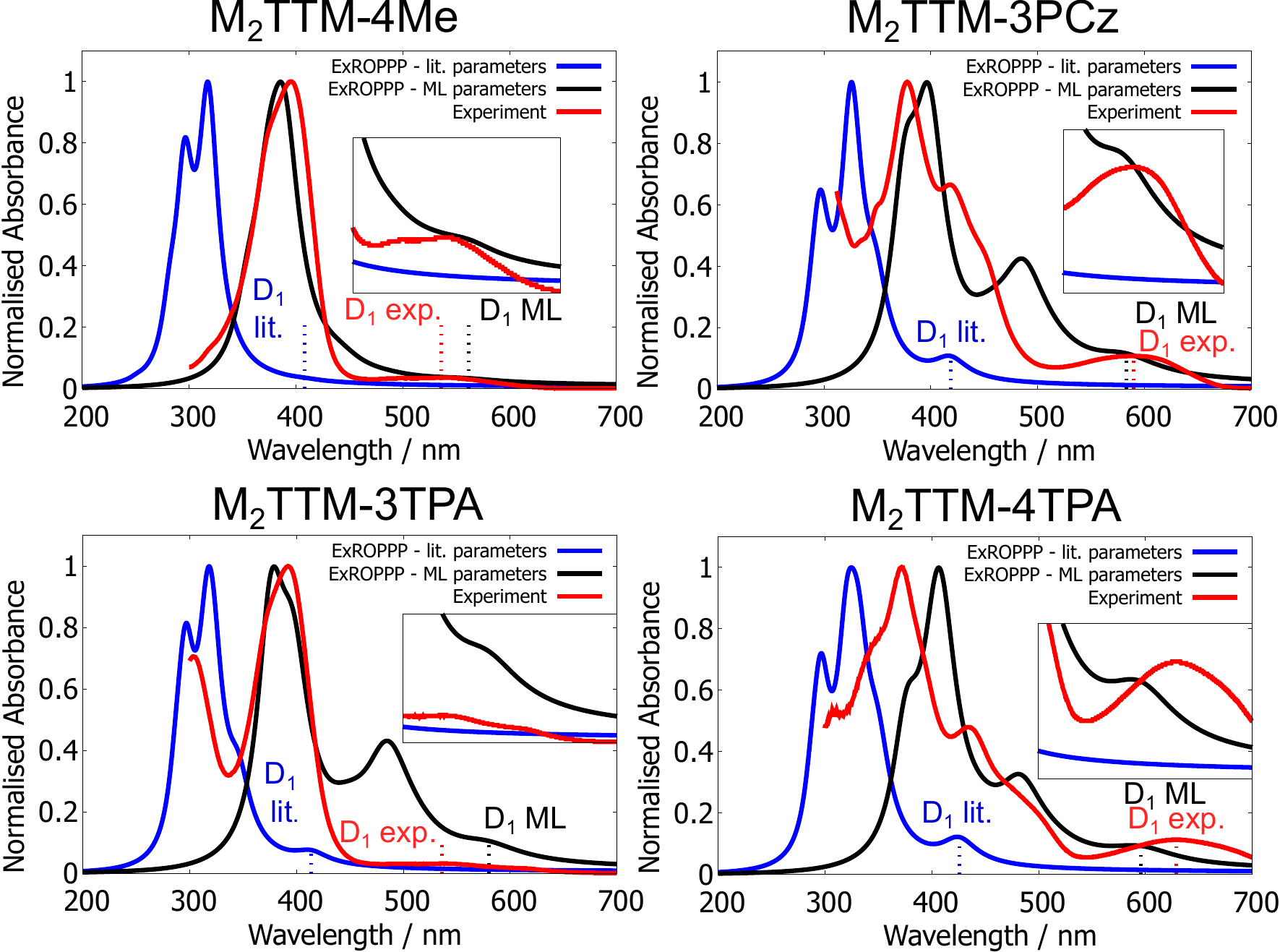}
\caption{UV-visible absorption spectra of the newly-synthesised M\textsubscript{2}TTM-4Me (top left), M\textsubscript{2}TTM-3PCz (top right), M\textsubscript{2}TTM-3TPA (bottom left) and M\textsubscript{2}TTM-4TPA (bottom right) measured in 0.1 mM toluene solution (red), simulated using ExROPPP with literature parameters (blue) and of the trained 81 molecule ExROPPP model (black). Trained ExROPPP (black) reproduces the experimental spectra (red) more accurately than untrained ExROPPP with literature parameters (blue).}
\figl{abs_test}
\end{figure*}

The synthesis of the radical species commenced with the formation of $\alpha$HM\textsubscript{2}TTM as previously reported by Murto \emph{et. al..}\cite{mur23a} Following this, $\alpha$HM\textsubscript{2}TTM was reacted with the respective 3- and 4- linked boronic acids of PCz and TPA to form $\alpha$HM\textsubscript{2}TTM-3PCz and $\alpha$HM\textsubscript{2}TTM-4TPA. To create the $\alpha$H precursors for the other two radical species, the remaining para-chlorine of $\alpha$HM\textsubscript{2}TTM was converted to a boronic ester through a Miyaura borylation before being coupled with 4-iodotoluene or 3-bromotriphenylamine. To convert into their respective radicals, all four $\alpha$H species were subjected to tetrabutylammonium hydroxide, to form the monoanion, before being oxidised to the radical using para-chloranil. M\textsubscript{2}TTM-4TPA, M\textsubscript{2}TTM-3TPA, M\textsubscript{2}TTM-3PCz and M\textsubscript{2}TTM-4T were formed in a 13\%, 56\%, 86\% and 37\% yield respectively.
UV-vis absorption measurements were carried out for the radicals in a 0.1 mM toluene solution. All four radicals display an intense absorption feature around 370-400 nm, which is characteristic of a local excitation within the TTM radical core. For M\textsubscript{2}TTM-3PCz and M\textsubscript{2}TTM-4TPA, additional absorption peaks can be seen at 590 and 630 nm respectively. These are attributed to CT transitions between the electron donating group and the electron-accepting TTM core.

We find that the four new molecules confirm the structure-property predictions made in 2020 \cite{abd20a} that, in order for a significant D$_1$ absorption the molecule should not be an alternant hydrocarbon and that the HOMO on the donor (4Me, 3PCz or TPA in this case) has orbital amplitude on the atom through which it is joined to the acceptor (TTM). M$_2$TTM-4Me is predicted to have minimal D$_1$ oscillator strength as it is a \emph{de facto} alternant hydrocarbon, as is observed experimentally. M$_2$TTM-3PCz contains a five-membered ring and a nitrogen, both of which break alternacy symmetry leading to a bright D$_1$ state, as is experimentally observed. A simple PPP calculation on TPA alone finds that the HOMO has significant amplitude at the para (4) position but minimal amplitude at the meta (3) position, as shown in \figr{tpa_homo}. This therefore predicts that the TPA to TTM charge transfer excitation will be dark in M$_2$TTM-3TPA but bright in M$_2$TTM-4TPA, as is observed experimentally in \figr{abs_test}.  We believe this is the first direct experimental confirmation of the design rule concerning the HOMO amplitude.

\subsection{Testing}
We tested the trained 81-molecule model on our four new organic radicals: M\textsubscript{2}TTM-4Me, M\textsubscript{2}TTM-3PCz, M\textsubscript{2}TTM-3TPA and M\textsubscript{2}TTM-4TPA shown in \figr{lucy_mol}, which make up the testing set. We find that the trained model performs well on the testing set, predicting both $D_1$ and bright state energies with a significantly higher accuracy than the literature parameters, as can be seen in \figr{comp_plot_test}. We also calculated the RMSE, MAD, $R^2$ and SRCC for the testing data, presented in \tabr{stats_test}. We find similar values for the errors and correlation metrics for the testing set as seen previously for the training set, again with RMSE and MAD less than 0.3 eV and $\mathrm{R}^2$ and SRCC of 0.93 and 0.76 respectively. The fact that the errors (RMSE and MAD) for the testing set are actually slightly lower than for the training set further indicates that overfitting did not occur. 

\begin{figure}[tb]
\centering
\includegraphics[width=0.45\textwidth]{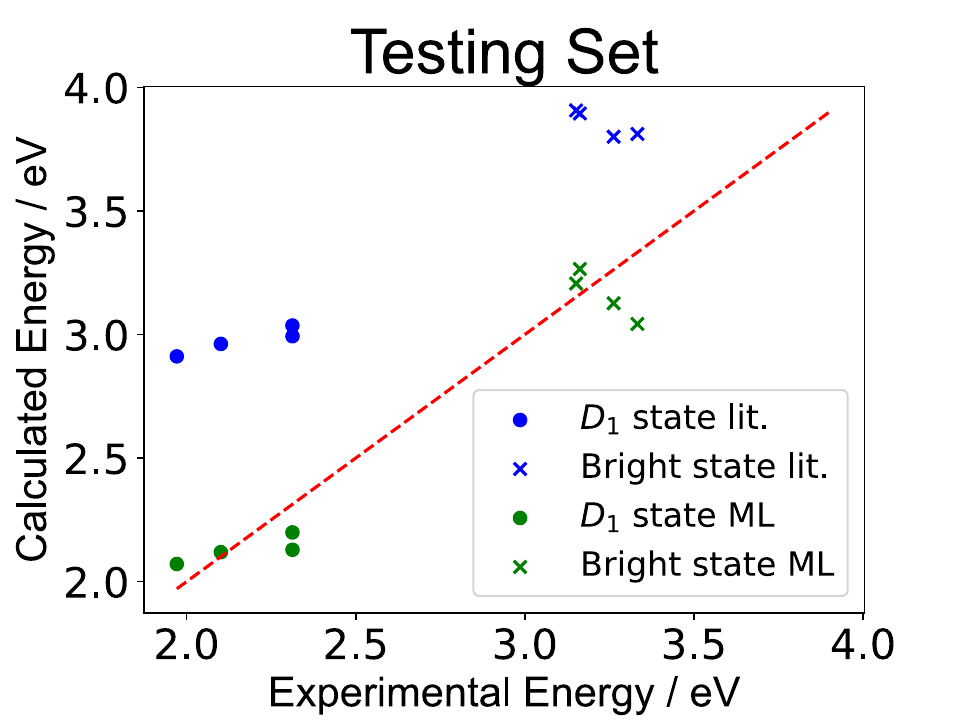}
\caption{Regression plot of the excited state energies of the radicals in the testing set calculated by ExROPPP and compared with experimentally determined energies. `lit.' refers to the parameters sourced from the literature and `ML' refers to the parameters of the trained 81 molecule model. It can be clearly seen that for this testing set the trained ExROPPP model more accurately reproduces the experimental values than ExROPPP with literature parameters.}
\figl{comp_plot_test}
\end{figure}

\begin{table}[tb]
\caption{Root mean-squared errors (RMSE), mean absolute differences (MAD), $\mathrm R^2$ and Spearman's rank correlation coefficients (SRCC) for the testing set of four newly synthesised molecules: M\textsubscript{2}TTM-4Me, M\textsubscript{2}TTM-3PCz, M\textsubscript{2}TTM-3TPA and M\textsubscript{2}TTM-4TPA, using i) the trained 81 molecule model and ii) ExROPPP with parameters obtained from the literature.}
\begin{tabular}{c|c|c|c}
&Literature& Trained  & Target\\ 
&parameters & Model & \\ \hline 
RMSE (all states) / eV & 0.73& 0.15& $<0.3$\\
MAD (all states) / eV & 0.71 & 0.13& $<0.3$\\   
$\mathrm R^2$ (all states) & -0.83&0.93 & close to 1 \\
SRCC (all states) & 0.76&0.76& close to 1
\end{tabular}
\tabl{stats_test}
\end{table}

We also compare the ExROPPP simulated UV-visible absorption spectra, with both literature and trained parameters, with the experimental spectra for these four molecules as shown in \figr{abs_test}. The simulated spectra of all four molecules are significantly improved after training. As well as a significant improvement in accuracy of the $D_1$ and bright state energies, the shape of the spectra are overall also better captured by the trained ExROPPP model. The only slight outlier is M\textsubscript{2}TTM-3TPA, for which ExROPPP predicts a larger $D_1$ intensity and lower $D_1$ energy than seen in experiment. Overall, however, the trained model accurately reproduces the absorption spectra of these four unseen molecules. 

\section{Conclusions}
\label{sec:con}
In this article we have presented the first demonstration of learning the excited states of radicals from experimental data. We achieve this by using the spin-pure ExROPPP method as a surrogate model, both to avoid the spin-contamination problem, and to address the limited experimental data in the literature. We find that the trained ExROPPP model performs far better at computing spectral features of organic radicals than the literature parameters. Four new radicals are synthesised and we test our model by comparing computed spectra against experimental data, finding good agreement and demonstrating its wider applicability as a predictive model. In future work this model could be further extended to predicting the emission spectra of radicals, and also to other atoms and groups common in organic radicals such as O, S and F, nitrile, nitro, aminoxyl and trifluoromethyl.\cite{miz24a}. In summary, this work serves as a major step forward for high-throughput screening and inverse molecular design of radicals with applications in OLEDs and qubits.

\section{Experimental Methods}
\subsection{Characterization techniques of organic radicals}
NMR spectra were acquired using a 400 MHz Bruker Avance III HD spectrometer (\textsuperscript{1}H, 400 MHz; \textsuperscript{13}C, 100 MHz). Chemical shifts are reported in $\delta$ (ppm) relative to the solvent peak: chloroform-d (CDCl\textsubscript{3}: \textsuperscript{1}H, 7.26 ppm; \textsuperscript{13}C, 77.16 ppm) and dichloromethane-d\textsubscript{2} (CD\textsubscript{2}Cl\textsubscript{2}: \textsuperscript{1}H, 5.32 ppm; \textsuperscript{13}C, 53.84 ppm). Mass spectra were obtained on a Waters Xevo G2-S benchtop QTOF mass spectrometer equipped with a electrospray ionization (ESI) or an atmospheric solid analysis probe (ASAP). Flash chromatography was carried out using Biotage Isolera Four System and Biotage SNAP/Sf\"ar Silica flash cartridges.

\subsection{Steady-state UV-visible spectroscopy}
UV–visible spectra were measured with a commercially available Shimadzu UV-1800 spectrophotometer.

\section*{Author Contributions}
TJHH conceived the research idea with assistance from JDG and JS. Computational research for training the ExROPPP model was undertaken by JDG, JS and KM, supervised by TJHH. Organic synthesis, characterization and measurement of UV-Visible absorption spectra was performed by LEW, supervised by HB. JS, KM, LEW and JDG all contributed to DFT calculations. The orbital visualiser was written by KM. KB advised on machine learning. The initial draft of the manuscript was written by JDG with assistance from TJHH, JS and KM and the organic synthetic parts written by LEW. All authors contributed to the revising and reviewing of the manuscript. 

\section*{Conflicts of interest}
There are no  conflicts to declare.

\section*{Data availability statement}
 %The Extended Data including the molecular geometries and literature UV-visible absorption data for all molecules, and initial and optimised sets of parameters were uploaded to the UCL Research Data Repository at https://rdr.ucl.ac.uk/...
 The Extended Data including the molecular geometries and literature UV-visible absorption data for all molecules, and initial and optimised sets of parameters is available at the UCL Research Data Repository at <link to be provided>
 %https://rdr.ucl.ac.uk/

\section*{Acknowledgments}
We thank Richard H.\ Friend for helpful discussions and Emrys W. Evans for providing geometries of TTM-3NCz, TTM-3PCz, TTM-$\alpha$PyID,TTM-$\beta$PyID, TTM-$\gamma$PyID and TTM-$\delta$PyID. TJHH acknowledges a Royal Society University Research Fellowship URF\textbackslash R1\textbackslash 201502 and a startup grant from University College London. HB acknowledges the grant EP/W017091/1. L.E.W. acknowledges funding from the European Research Council under the European Union’s Horizon 2020 research and innovation program grant agreement No. 101020167. KTB acknowledges funding from UKRI grants EP/Y000552/1 and EP/Y014405/1. We thank Feng Li for permission to reuse the structure of CzBTM.

\bibliography{exroppp-ml}

\begin{thebibliography}{10}

\bibitem{pen15a}
Q.~Peng, A.~Obolda, M.~Zhang and F.~Li, \emph{Angew. Chem., Int. Ed.} \textbf{54} (2015), 7091.

\bibitem{ai18a}
X.~Ai, E.~W. Evans, S.~Dong, A.~J. Gillett, H.~Guo, Y.~Chen, T.~J.~H. Hele, R.~H. Friend and F.~Li, \emph{Nature} \textbf{563} (2018), 536.

\bibitem{guo14a}
Z.~Guo, S.~Park, J.~Yoon and I.~Shin, \emph{Chemical Society Reviews} \textbf{43} (2014), 16.

\bibitem{bla02a}
M.~Blanco and I.~Villarroya, \emph{TrAC Trends in Analytical Chemistry} \textbf{21} (2002), 240.

\bibitem{mur23a}
P.~Murto, R.~Chowdhury, S.~Gorgon, E.~Guo, W.~Zeng, B.~Li, Y.~Sun, H.~Francis, R.~H. Friend and H.~Bronstein, \emph{Nature Communications} \textbf{14} (2023), 4147.

\bibitem{abd20a}
A.~Abdurahman, T.~J. Hele, Q.~Gu, J.~Zhang, Q.~Peng, M.~Zhang, R.~H. Friend, F.~Li and E.~W. Evans, \emph{Nature materials} \textbf{19} (2020), 1224.

\bibitem{li22a}
F.~Li, A.~J. Gillett, Q.~Gu, J.~Ding, Z.~Chen, T.~J. Hele, W.~K. Myers, R.~H. Friend and E.~W. Evans, \emph{Nature Communications} \textbf{13} (2022), 2744.

\bibitem{cho24a}
H.-H. Cho, S.~Gorgon, G.~Londi, S.~Giannini, C.~Cho, P.~Ghosh, C.~Tonnel{\'e}, D.~Casanova, Y.~Olivier, T.~K. Baikie \emph{et~al.}, \emph{Nature Photonics}  (2024), 1.

\bibitem{gor23a}
S.~Gorgon, K.~Lv, J.~Gr{\"u}ne, B.~H. Drummond, W.~K. Myers, G.~Londi, G.~Ricci, D.~Valverde, C.~Tonnel{\'e}, P.~Murto, A.~S. Romanov, D.~Casanova, V.~Dyakonov, A.~Sperlich, D.~Beljonne, Y.~Olivier, F.~Li, R.~H. Friend and E.~W. Evans, \emph{Nature} \textbf{620} (2023), 538.

\bibitem{yu24a}
C.~P. Yu, R.~Chowdhury, Y.~Fu, P.~Ghosh, W.~Zeng, T.~B. Mustafa, J.~Gr{\"u}ne, L.~E. Walker, D.~G. Congrave, X.~W. Chua \emph{et~al.}, \emph{Science Advances} \textbf{10} (2024), eado3476.

\bibitem{miz24b}
A.~Mizuno, R.~Matsuoka, S.~Kimura, K.~Ochiai and T.~Kusamoto, \emph{Journal of the American Chemical Society} \textbf{146} (2024), 18470, pMID: 38921686.

\bibitem{and92a}
K.~Andersson, P.-{\AA}. Malmqvist and B.~O. Roos, \emph{The Journal of chemical physics} \textbf{96} (1992), 1218.

\bibitem{nak93a}
H.~Nakano, \emph{The Journal of Chemical Physics} \textbf{99} (1993), 7983.

\bibitem{pur82a}
G.~D. Purvis~III and R.~J. Bartlett, \emph{The Journal of chemical physics} \textbf{76} (1982), 1910.

\bibitem{he19a}
C.~He, Z.~Li, Y.~Lei, W.~Zou and B.~Suo, \emph{The Journal of Physical Chemistry Letters} \textbf{10} (2019), 574.

\bibitem{li16a}
Z.~Li and W.~Liu, \emph{Journal of Chemical Theory and Computation} \textbf{12} (2016), 238, pMID: 26672389.

\bibitem{hel21b}
T.~J.~H. Hele, B.~Monserrat and A.~M. Alvertis, \emph{The Journal of Chemical Physics} \textbf{154} (2021), 244109.

\bibitem{gre24a}
J.~D. Green and T.~J.~H. Hele, \emph{The Journal of Chemical Physics} \textbf{160} (2024), 164110.

\bibitem{par53a}
R.~Pariser and R.~G. Parr, \emph{The Journal of Chemical Physics} \textbf{21} (1953), 466.

\bibitem{par56a}
R.~Pariser, \emph{The Journal of Chemical Physics} \textbf{24} (1956), 250.

\bibitem{pop53a}
J.~A. Pople, \emph{Trans. Faraday Soc.} \textbf{49} (1953), 1375.

\bibitem{pop54a}
J.~A. Pople and R.~K. Nesbet, \emph{The Journal of Chemical Physics} \textbf{22} (1954), 571.

\bibitem{mat57a}
N.~Mataga and K.~Nishimoto, \emph{Zeitschrift f{\"u}r Physikalische Chemie} \textbf{13} (1957), 140.

\bibitem{mau96a}
D.~Maurice and M.~Head-Gordon, \emph{The Journal of Physical Chemistry} \textbf{100} (1996), 6131.

\bibitem{hel19a}
T.~J.~H. Hele, E.~G. Fuemmeler, S.~N. Sanders, E.~Kumarasamy, M.~Y. Sfeir, L.~M. Campos and N.~Ananth, \emph{The Journal of Physical Chemistry A} \textbf{123} (2019), 2527, pMID: 30802051.

\bibitem{gre22a}
J.~D. Green, E.~G. Fuemmeler and T.~J.~H. Hele, \emph{The Journal of Chemical Physics} \textbf{156} (2022), 180901.

\bibitem{bed23a}
M.~Bedogni, D.~Giavazzi, F.~Di~Maiolo and A.~Painelli, \emph{Journal of Chemical Theory and Computation} \textbf{20} (2023), 902.

\bibitem{dub24a}
M.~Dubbini, F.~Bonvini, L.~Savi and F.~Di~Maiolo, \emph{The Journal of Physical Chemistry C} \textbf{128} (2024), 18158.

\bibitem{jor24a}
K.~Jorner, R.~Pollice, C.~Lavigne and A.~Aspuru-Guzik, \emph{The Journal of Physical Chemistry A} \textbf{128} (2024), 2445, pMID: 38485448.

\bibitem{hin71a}
J.~Hinze and D.~L. Beveridge, \emph{Journal of the American Chemical Society} \textbf{93} (1971), 3107.

\bibitem{van80a}
F.~Van-Catledge, \emph{The Journal of Organic Chemistry} \textbf{45} (1980), 4801.

\bibitem{but18a}
K.~T. Butler, D.~W. Davies, H.~Cartwright, O.~Isayev and A.~Walsh, \emph{Nature} \textbf{559} (2018), 547.

\bibitem{mon13a}
G.~Montavon, M.~Rupp, V.~Gobre, A.~Vazquez-Mayagoitia, K.~Hansen, A.~Tkatchenko, K.-R. M{\"u}ller and O.~A. Von~Lilienfeld, \emph{New Journal of Physics} \textbf{15} (2013), 095003.

\bibitem{ben18a}
B.~Sanchez-Lengeling and A.~Aspuru-Guzik, \emph{Science} \textbf{361} (2018), 360.

\bibitem{bar17a}
A.~P. Bartók, S.~De, C.~Poelking, N.~Bernstein, J.~R. Kermode, G.~Csányi and M.~Ceriotti, \emph{Science Advances} \textbf{3} (2017), e1701816.

\bibitem{der19a}
V.~L. Deringer, M.~A. Caro and G.~Csányi, \emph{Advanced Materials} \textbf{31} (2019), 1902765.

\bibitem{srs24a}
{\v{S}}.~Sr{\v{s}}e{\v{n}}, O.~A. von Lilienfeld and P.~Slaví{\v{c}}ek, \emph{Phys. Chem. Chem. Phys.} \textbf{26} (2024), 4306.

\bibitem{che18a}
W.-K. Chen, X.-Y. Liu, W.-H. Fang, P.~O. Dral and G.~Cui, \emph{The Journal of Physical Chemistry Letters} \textbf{9} (2018), 6702.

\bibitem{ram15a}
R.~Ramakrishnan, M.~Hartmann, E.~Tapavicza and O.~A. von Lilienfeld, \emph{The Journal of Chemical Physics} \textbf{143} (2015), 084111.

\bibitem{roc20a}
L.~M. Roch, S.~K. Saikin, F.~Häse, P.~Friederich, R.~H. Goldsmith, S.~León and A.~Aspuru-Guzik, \emph{ACS Nano} \textbf{14} (2020), 6589, pMID: 32338888.

\bibitem{che22b}
Z.~Chen, F.~C. Bononi, C.~A. Sievers, W.-Y. Kong and D.~Donadio, \emph{Journal of Chemical Theory and Computation} \textbf{18} (2022), 4891, pMID: 35913220.

\bibitem{wes21a}
J.~Westermayr and P.~Marquetand, \emph{Chemical Reviews} \textbf{121} (2021), 9873, pMID: 33211478.

\bibitem{dra21a}
P.~O. Dral and M.~Barbatti, \emph{Nature Reviews Chemistry} \textbf{5} (2021), 388.

\bibitem{dra18a}
P.~O. Dral, M.~Barbatti and W.~Thiel, \emph{The journal of physical chemistry letters} \textbf{9} (2018), 5660.

\bibitem{xue20a}
B.-X. Xue, M.~Barbatti and P.~O. Dral, \emph{The Journal of Physical Chemistry A} \textbf{124} (2020), 7199, pMID: 32786977.

\bibitem{ju24a}
C.-W. Ju, Y.~Shen, E.~J. French, J.~Yi, H.~Bi, A.~Tian and Z.~Lin, \emph{The Journal of Physical Chemistry A} \textbf{128} (2024), 2457, pMID: 38382058.

\bibitem{fab22a}
A.~Fabrizio, K.~R. Briling and C.~Corminboeuf, \emph{Digital Discovery} \textbf{1} (2022), 286.

\bibitem{bri24a}
K.~R. Briling, Y.~Calvino~Alonso, A.~Fabrizio and C.~Corminboeuf, \emph{Journal of Chemical Theory and Computation} \textbf{20} (2024), 1108, pMID: 38227222.

\bibitem{wes20a}
J.~Westermayr, M.~Gastegger and P.~Marquetand, \emph{The Journal of Physical Chemistry Letters} \textbf{11} (2020), 3828, pMID: 32311258.

\bibitem{hof63a}
R.~Hoffmann, \emph{The Journal of Chemical Physics} \textbf{39} (1963), 1397.

\bibitem{dew77a}
M.~J. Dewar and W.~Thiel, \emph{Journal of the American Chemical Society} \textbf{99} (1977), 4899.

\bibitem{ste07a}
J.~J. Stewart, \emph{Journal of Molecular modeling} \textbf{13} (2007), 1173.

\bibitem{rep02a}
M.~P. Repasky, J.~Chandrasekhar and W.~L. Jorgensen, \emph{Journal of computational chemistry} \textbf{23} (2002), 1601.

\bibitem{chr16a}
A.~S. Christensen, T.~Kubař, Q.~Cui and M.~Elstner, \emph{Chemical Reviews} \textbf{116} (2016), 5301, pMID: 27074247.

\bibitem{fab08a}
E.~Fabiano, T.~Keal and W.~Thiel, \emph{Chemical Physics} \textbf{349} (2008), 334, electron Correlation and Molecular Dynamics for Excited States and Photochemistry.

\bibitem{hel21a}
T.~J.~H. Hele, \emph{Physical Chemistry of Semiconductor Materials and Interfaces XX}, International Society for Optics and Photonics, SPIE (2021), volume 11799, 117991A.

\bibitem{chu54a}
T.~Li~Chu and S.~I. Weissman, \emph{The Journal of Chemical Physics} \textbf{22} (1954), 21.

\bibitem{lon55a}
H.~C. Longuet-Higgins and J.~A. Pople, \emph{Proceedings of the Physical Society. Section A} \textbf{68} (1955), 591.

\bibitem{hud21a}
J.~M. Hudson, T.~J.~H. Hele and E.~W. Evans, \emph{Journal of Applied Physics} \textbf{129} (2021), 180901.

\bibitem{miz24a}
A.~Mizuno, R.~Matsuoka, T.~Mibu and T.~Kusamoto, \emph{Chemical Reviews} \textbf{124} (2024), 1034, pMID: 38230673.

\bibitem{ai18b}
X.~Ai, Y.~Chen, Y.~Feng and F.~Li, \emph{Angewandte Chemie International Edition} \textbf{57} (2018), 2869.

\bibitem{rub93a}
S.~R. Ruberu and M.~A. Fox, \emph{The Journal of Physical Chemistry} \textbf{97} (1993), 143.

\bibitem{mat22a}
K.~Matsuda, R.~Xiaotian, K.~Nakamura, M.~Furukori, T.~Hosokai, K.~Anraku, K.~Nakao and K.~Albrecht, \emph{Chem. Commun.} \textbf{58} (2022), 13443.

\bibitem{gao17a}
Y.~Gao, A.~Obolda, M.~Zhang and F.~Li, \emph{Dyes and Pigments} \textbf{139} (2017), 644.

\bibitem{don17a}
S.~Dong, A.~Obolda, Q.~Peng, Y.~Zhang, S.~Marder and F.~Li, \emph{Mater. Chem. Front.} \textbf{1} (2017), 2132.

\bibitem{guo19a}
H.~Guo, Q.~Peng, X.-K. Chen, Q.~Gu, S.~Dong, E.~W. Evans, A.~J. Gillett, X.~Ai, M.~Zhang, D.~Credgington \emph{et~al.}, \emph{Nature materials} \textbf{18} (2019), 977.

\bibitem{don18a}
S.~Dong, W.~Xu, H.~Guo, W.~Yan, M.~Zhang and F.~Li, \emph{Phys. Chem. Chem. Phys.} \textbf{20} (2018), 18657.

\bibitem{yan22a}
C.~Yan, D.~An, W.~Chen, N.~Zhang, Y.~Qiao, J.~Fang, X.~Lu, G.~Zhou and Y.~Liu, \emph{CCS Chemistry} \textbf{4} (2022), 3190.

\bibitem{gamess}
G.~M.~J. Barca, C.~Bertoni, L.~Carrington, D.~Datta, N.~De~Silva, J.~E. Deustua, D.~G. Fedorov, J.~R. Gour, A.~O. Gunina, E.~Guidez, T.~Harville, S.~Irle, J.~Ivanic, K.~Kowalski, S.~S. Leang, H.~Li, W.~Li, J.~J. Lutz, I.~Magoulas, J.~Mato, V.~Mironov, H.~Nakata, B.~Q. Pham, P.~Piecuch, D.~Poole, S.~R. Pruitt, A.~P. Rendell, L.~B. Roskop, K.~Ruedenberg, T.~Sattasathuchana, M.~W. Schmidt, J.~Shen, L.~Slipchenko, M.~Sosonkina, V.~Sundriyal, A.~Tiwari, J.~L. Galvez~Vallejo, B.~Westheimer, M.~Włoch, P.~Xu, F.~Zahariev and M.~S. Gordon, \emph{The Journal of Chemical Physics} \textbf{152} (2020), 154102.

\bibitem{sei21a}
N.~A. Seifert, K.~Prozument and M.~J. Davis, \emph{The Journal of Chemical Physics} \textbf{155} (2021).

\bibitem{nel65a}
J.~A. Nelder and R.~Mead, \emph{The Computer Journal} \textbf{7} (1965), 308.

\bibitem{bru71a}
S.~{de Bruijn}, \emph{Chemical Physics Letters} \textbf{8} (1971), 19.

\bibitem{gho24a}
P.~Ghosh, A.~M. Alvertis, R.~Chowdhury, P.~Murto, A.~J. Gillett, S.~Dong, A.~J. Sneyd, H.-H. Cho, E.~W. Evans, B.~Monserrat \emph{et~al.}, \emph{Nature} \textbf{629} (2024), 355.

\end{thebibliography}

%\section*{to-do}
%\begin{itemize}
%\item{references here and there}
%\item people in EPR community are interested in calculating radicals for spin-density etc but not excited states
%\item chemdraw structures for lucy molecules (jingkun)
%\item Side by side comparison of ExROPPP vs ROHF orbitals
%\item side by side plots of spectra for training molecules
%\item{Some more suggestions for futw are: could instead train ExROPPP on EOM-CCSD spectra (PySCF library has inbuilt EOM-CCSD code, could optimize geometries using PySCF too such that the workflow is more automated. Don't need to go through many papers and pick out spectra, optimised geometries automatically added to dataset to save time/confusion), could investigate alternative fitness functions on ExROPPP training process such as optimal transport}
%\end{itemize}
\end{document}